\documentclass[12pt]{article}

\usepackage{amsfonts}

\newcommand{\beq}{\begin{equation}}
\newcommand{\eeq}{\end{equation}}
\newcommand{\bea}{\begin{eqnarray}}
\newcommand{\eea}{\end{eqnarray}}

\begin{document}

\begin{center}
{\Large {\bf DEDUCTION OF PLANCK'S FORMULA FROM MULTIPHOTON STATES}}
\end{center}
\hfil\break
\begin{center}
{\bf Luis J. Boya}\footnote{Permanent Address: Departmento de F\'\i sica 
Te\'orica, Facultad de Ciencias.  Universidad de Zaragoza. E-50009 Zaragoza, Spain.  
E-mail: luisjo@posta.unizar.es}, 
{\bf Ian M. Duck}\footnote{Physics Department, Rice University, Houston, TX. 77251-1892 }
 and 
{\bf E. C. G. Sudarshan} \footnote{E-mail: sudarshan@utaphy.ph.utexas.edu}

Center for Particle Physics\\
Department of Physics\\
The University of Texas at Austin\\
Austin, Texas 78712-1081, USA
\end{center}

\begin{flushleft}
{\it Key Words:}  photons, radiation law, old quantum theory.
\end{flushleft}

\begin{abstract}
We obtain the black body radiation formula of Planck by considering independent 
contributions of multiphoton entities
\end{abstract}

\vfill
\pagebreak


{\Large{\bf 1.}}  Einstein introduced the light quantum hypothesis in 1905 \cite{1} 
by noticing that the entropy calculated from the Wien formula for the black body radiation

\begin{equation}
	I_W(\nu,T) = a \nu^3 \exp(-b\nu/T),
\end{equation}
which is the high-frequency, low-density limit of the complete Planck's formula
\begin{equation}
		I(\nu, T) = \frac{8{\pi}h{\nu}^3}{c^3}  \frac{1}{e^{(h{\nu}/kT)} -1},
\end{equation}
this entropy had the same form as the entropy of a ``gas'' of independent particles, 
{\it light quanta} (later called photons). Viceversa, starting from a gas of independent 
light quanta one arrives only to Wien's formula, as shown since 1910 by many 
people \cite{6}.
	
The question arises as how to get the complete radiation formula from the 
corpuscular concept of the photon. In this letter we present what we think is about 
the simplest derivation of Planck's radiation formula, starting from the idea of 
``multiphoton states'' \cite{2}, i.e. molecules of light compound of $n$ photons with 
zero binding energy.
	
We know of course that the most satisfactory deduction of Planck's formula 
is obtained nowadays by incorporating the right B.E. statistics into the corpuscular 
picture, as first shown by S.N.Bose \cite{3} and interpreted and extended by Einstein 
in 1924/25 \cite{4}. Our calculation makes use of well-defined multiphoton states as 
the natural way of taking in account the identical, indistinguishable character 
of the photons.
	
The idea of multiphoton ``molecules'' arose around 1910, after P. Debye 
presented \cite{5} a simple, ondulatory method to obtain the radiation law;  some 
historical comments are included below in this letter.

\bigskip
{\Large{\bf 2.}}  We start from the photon  entity as particles of energy 
${\epsilon}=h{\nu}$ (they are truly massless particles of helicity $\pm 1$
with energy and momentum ${\epsilon}=h{\nu}=pc, p$, but we want to 
think and write in the spirit of the 1910s).
	
We make now this additional simple hypothesis \cite{2}: {\it light} of definite 
direction and frequency ${\nu}$ presents itself in {\it units} (``molecules'') 
of $0,1,2,...,n,... $photons, with energy $nh{\nu}$, that is, with zero binding 
energy, and they contribute {\it independently} to the energy density.

	The rationale for this assumption is, of course, in the spirit of the atomistic
 or democritean point of view, that all the photons of same frequency are created equal
 (identical),
 and that in an occamian, economical description, we should abstain from considering
 situations in which
the photons are individually distinct (indistinguishability); that is, there is a 
``state'' with two photons {\it different}, for the counting, from the naive 
juxtaposition of two
distinguishable photons; same thing for the three photon states, etc. So the 
multiphoton molecules should be thought of as entities only to the effect of 
counting, and not as
claiming for the existence of some force which really binds the photons together.
	
	In this way one gets the correct radiation formula at once. We are 
unaware of any complete proof of this statement, so  we now show the deduction, 
leaving some historical remarks for later.
	
For the $n$-photon  `` molecules''  of energy $nh{\nu },(n=0,1,2,...)$ the probability
 is of course given by the Boltzmann factor 
\begin{equation}
p(n)=c\; \exp(-nh{\nu}/kT) 
\end{equation}
\begin{equation}
	\sum_{n=0}^\infty p(n) =1 \;\mbox{gives}\; c=1-x,\;\; x \equiv \exp(-h{\nu}/kT).    
\end{equation}

	The multiphoton states still have frequency ${\nu}$, so the common phase 
space factor is $2$(polarization)$\times 4{\pi}$(solid angle)$\times {\nu^2}$ 
(radial 3D factor);
 hence the density of energy per inverse wavelength is 
\begin{eqnarray}
I({\nu}, T) &=& 8 \pi ( \nu^2 /c^3)\Sigma_0^{\infty} nh \nu(1-x)\exp(-nh \nu/kT)\nonumber\\
            &=& \frac{8\pi \nu^3}{c^3} \frac{hx}{1-x} \nonumber \\ 
            &=& \frac{8\pi\nu^3h}{c^3} \frac{1}{\exp(h{\nu}/kT) -1}
\end{eqnarray}
i.e., the correct complete radiation formula.
	
The two traditional limits of the radiation formula do have a particle interpretation,
 as already remarked by Wolfke [2]. E.g. at low temperature only the lightest 
``molecule'' is excited, that is, the one-photon mode, of energy $h{\nu}$, that leads 
to the Wien's limit law (1), which was Einstein's starting point \cite{1}.

	On the other hand, for high densities or temperatures all the $n$-photon states
contribute together and sort of coalesce, giving a scale-invariant power law distribution
$I \propto {\nu^2}$, i.e. the classical electromagnetic (Rayleigh-Jeans)limit formula.

\bigskip
{\Large {\bf 3.}}  If we now start from (5) and expand the denominator, we get 
\begin{equation}
	I({\nu},T) = 8\pi({\nu}^2/c^3) (h{\nu}) \sum_{n=1}^{\infty} \exp(-nh{\nu}/kT)  
\end{equation}
already obtained by several people \cite{6} in the 1910s; superficially, (6) looks like an
 addditive contribution from each $n$-photon state, and it was this analogy which 
prompted Ehrenfest \cite{7}, Wolfke \cite{2}, de Broglie \cite{8,9} and others \cite{6} 
to advance the ``$n$-photon molecule'' concept; notice however that the {\it literal} 
interpretation of (6) as a cooperative multiphoton formula falls short of the correct 
derivation in (5) on three accounts:
\begin{enumerate}	
 \item There is no contribution from the zero-photon state
 \item The $n-$photon state contributes with energy $h{\nu}$, instead of the
correct $nh{\nu}$
 \item Probabilities $\exp(-nh{\nu}/kT)$ do {\it not} add up to one. 
\end{enumerate}
Still, it is remarkable that the expansions (5) and (6) are really identical!

\bigskip
{\Large {\bf 4.}} Why our simple derivation was not, to the best of our knowledge, obtained
 and exploited many decades ago, is for the historians to discuss; perhaps light quanta
 correlations due to indiscernibility was too novel at the time. Recall, however, that the
Gibbs paradox was well known at the time, with the proposed solution by Gibbs himself [10].
	Notice that this treatment of photons is totally correct even from the modern point
 of view; in particular the zero-point energy does not show up. Also we are really doing 
second quantization. Namely ``first quantization'' just amounts to counting frequency 
(or wavenumber) modes, because ${\nu}=pc/h$; the ``second'' quantization amounts to 
the existence of these photon molecules as separate entities. We finish by remarking 
that zero-energy
 bound states appear frequently in modern guise as Bogomol'nyi-Prasad-Sommerfield (BPS) 
states, and moreover uncharged PBS states, like the photon, are massless.
\pagebreak


\end{document}